\documentclass[10pt,conference]{IEEEtran}
\IEEEoverridecommandlockouts
\usepackage{cite}
\usepackage{amsmath,amssymb,amsfonts}
\usepackage{algorithmic}
\usepackage{graphicx}
\usepackage{comment}
\usepackage{textcomp}
\usepackage{xcolor}
\def\BibTeX{{\rm B\kern-.05em{\sc i\kern-.025em b}\kern-.08em
    T\kern-.1667em\lower.7ex\hbox{E}\kern-.125emX}}
\usepackage[flushleft]{threeparttable}
\usepackage{pifont}
\usepackage{framed}
\usepackage{soul}
\usepackage{booktabs}  
\usepackage{csquotes}
\usepackage{multirow}
\usepackage{array}
\usepackage{balance}
\newcolumntype{L}[1]{>{\raggedright\let\newline\\\arraybackslash\hspace{0pt}}m{#1}}
\newcolumntype{C}[1]{>{\centering\let\newline\\\arraybackslash\hspace{0pt}}m{#1}}
\newcolumntype{R}[1]{>{\raggedleft\let\newline\\\arraybackslash\hspace{0pt}}m{#1}}
\usepackage{url}
\MakeOuterQuote{"}
\usepackage[caption=false,font=footnotesize]{subfig} 

\usepackage[bookmarks=false]{hyperref}

\usepackage{acronym}
\acrodef{IP}[IP]{intellectual property block}
\acrodef{SoC}[SoC]{System-on-Chip}
\acrodef{IC}[IC]{integrated circuit}
\acrodef{eFPGA}[eFPGA]{embedded field programmable gate array}
\acrodef{FPGA}[FPGA]{field programmable gate array}
\acrodef{RTL}[RTL]{register transfer level}
\acrodef{CLB}[CLB]{configurable logic block}
\acrodef{LUT}[LUT]{look-up table}
\acrodef{HLS}[HLS]{high-level synthesis}
\acrodef{EDA}[EDA]{electronic design automation}
\acrodef{FF}[FF]{flip-flop}
\acrodef{DIP}[DIP]{distinguishing input pattern}

\author{%
    \IEEEauthorblockN{Jitendra Bhandari\IEEEauthorrefmark{1}, Abdul Khader Thalakkattu Moosa\IEEEauthorrefmark{1}, Benjamin Tan\IEEEauthorrefmark{1}, Christian Pilato\IEEEauthorrefmark{2},\\ Ganesh Gore\IEEEauthorrefmark{3}, Xifan Tang\IEEEauthorrefmark{3}, Scott Temple\IEEEauthorrefmark{3}, Pierre-Emmanuel Gaillardon\IEEEauthorrefmark{3} and Ramesh Karri\IEEEauthorrefmark{1}\thanks{Benjamin Tan and Ramesh Karri are supported in part by the Office of Naval Research under Award Number \#N00014-18-1-2058. This work was supported in part by NYU CCS. Ganesh Gore, Xifan Tang, Pierre-Emmanuel Gaillardon are supported by AFRL and DARPA under agreement number FA8650-18-2-7855, and Scott Temple, Pierre-Emmanuel Gaillardon are supported by AFRL and DARPA under agreement number FA8650-18-2-7849. Jitendra Bhandari and Abdul Khader Thalakkattu Moosa contributed equally to this work.}}
    \IEEEauthorblockA{%
    \IEEEauthorrefmark{1}New York University, New York, USA\\
    \IEEEauthorrefmark{2}Politecnico di Milano, Milan, Italy \\
    \IEEEauthorrefmark{3}University of Utah, Utah, USA}

}

\begin{document}

\bstctlcite{IEEEexample:BSTcontrol}


\title{Exploring eFPGA-based Redaction for IP Protection}

\maketitle

\begin{abstract}
Recently, eFPGA-based redaction has been proposed as a promising solution for hiding parts of a digital design from untrusted entities, where legitimate end-users can restore functionality by loading the withheld bitstream after fabrication. 
However, when deciding which parts of a design to redact, there are a number of practical issues that designers need to consider, including area and timing overheads, as well as security factors. 
Adapting an open-source FPGA fabric generation flow, we perform a case study to explore the trade-offs when redacting different modules of open-source intellectual property blocks (IPs) and explore how different parts of an eFPGA contribute to the security. 
We provide new insights into the feasibility and challenges of using eFPGA-based redaction as a security solution. 
\end{abstract}



\section{Introduction}

\begin{figure}[t]
    \centering
    \includegraphics[width=0.7\columnwidth]{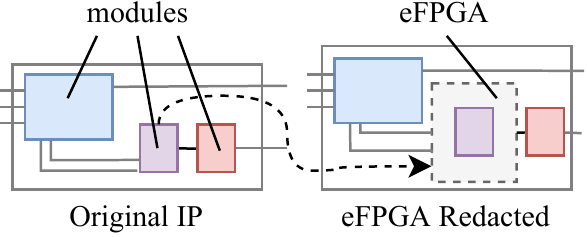}
    \caption{eFPGA-based redaction takes a module of an IP and replaces it with a reprogrammable fabric that can replace the redacted functionality. 
    }
    \label{fig:overview}
\end{figure}

In response to concerns about the integrity of the \ac{IC} supply-chain, researchers have proposed numerous design obfuscation and locking techniques to protect hardware \acp{IP}~\cite{pilato_assure_2021,LLC,shamsi_ip_2019,mohan_hardware_2021, chen_decoy_2020, hu_functional_2019, kamali_interlock_2020,kolhe_security_2019, liu_embedded_2014,limaye_thwarting_2020}. 
Such techniques involve supplementing designs so as to induce errors in the presence of incorrect key inputs (e.g., adding XOR/XNOR gates randomly~\cite{limaye_thwarting_2020}) or introducing structures that legitimate users later populate with elements of the design that are withheld during fabrication (e.g., restoring withheld constants~\cite{pilato_assure_2021}). 
Programmable fabrics have been added to the repertoire of defenses against reverse engineering and \ac{IP} piracy, especially as a counter to Boolean satisfiability-based (SAT) attacks~\cite{subramanyan_evaluating_2015} and variants thereof~\cite{LLC}.
In \textit{reconfigurable fabric-based redaction}, designers select parts of a design and implement it by programming a fabric separate from the remaining design, as shown in~\autoref{fig:overview}.  A potentially un-trusted foundry manufactures the design without the programming information for the fabric (e.g., the configuration bit-stream) which the designer withholds. Fabrics include \acused{FPGA}\acp{eFPGA}~\cite{hu_functional_2019,mohan_hardware_2021}, coarse-grained reconfigurable architectures~\cite{chen_area_2021}, and transistor  fabrics~\cite{shihab_design_2019}. 

When adopting \ac{eFPGA}-based redaction for \ac{IP} protection, designers are faced with a number of decisions and design challenges. 
These include customizing/selecting the fabric configuration, deciding which of the module(s) in the \ac{IP} to move to the \ac{eFPGA}, and dealing with the overheads that result from the process of integration and implementation in the ASIC design flow. 
Prior work has begun to explore these challenges by selecting functionality to redact from high-level (C-based) designs to maximize a security metric until hitting an area overhead threshold~\cite{hu_functional_2019} or by choosing a single part of a design---at the \ac{RTL}---which guides the generation of an \ac{eFPGA} fabric~\cite{mohan_hardware_2021}. 
As with many security solutions, there is a trade-off between security and other design factors, such as area and timing~\cite{chen_decoy_2020,hu_functional_2019,mohan_hardware_2021}. 
Thus far, prior work has suggested \acp{eFPGA} for feasibly redacting an individual \ac{IP}~\cite{hu_functional_2019,mohan_hardware_2021}; however, given the nascent state of \ac{eFPGA}-based redaction, we need more insights into the practical implications of using this \ac{IP} protection approach so as to help designers make better redaction decisions. 

In this work, we provide key insights into the use of \acp{eFPGA} for redacting \ac{RTL} designs through a case study of redacting different hand-crafted hierarchical \ac{RTL} \acp{IP}, where different modules are candidate units for redaction. 
We investigate adapting an open-source \acs{FPGA} design flow~\cite{tang_openfpga_2020} to produce different \acp{eFPGA} configurations, depending on the module to be redacted, and assess the impact on a range of open-source \acp{IP}. We explore factors that contribute to the security offered by \acp{eFPGA}-based redaction, and explore the factors that contribute to the security of eFPGA fabrics. 
The three main contributions are:
\begin{itemize}
    \item insights from a security evaluation of \ac{eFPGA}-based redaction based on different redaction decisions, under an oracle-based, scan-chain accessible attack model 
    \item results from a case study using open-source \acp{IP} to explore challenges and area/timing trade-offs of different redaction decisions, using an open-source \ac{FPGA} tool flow~\cite{tang_openfpga_2020} 
    \item insights and guidance for future work in \ac{eFPGA}-based redaction of \ac{RTL} designs
\end{itemize}
%
In \autoref{sec:2_prelims}, we provide the context of our work and describe the potential of open-source \acp{eFPGA} for hardware security. 
In \autoref{sec:proposed}, we provide an overview of an \ac{eFPGA}-based redaction flow, with a particular focus on the decision and challenges faced by designers. 
\autoref{sec:security} is our deep dive into the characteristics of \ac{eFPGA} fabrics that provide security benefits of redaction. 
We present the impact of different module choices using a set of open-source \acp{IP} from locking/obfuscation work, in \autoref{sec:results}, and discuss insights from our study in \autoref{sec:discussion}. 

\section{Motivation and Background \label{sec:2_prelims}}

\subsection{Related Work: Intellectual Property Protection}


Traditional techniques for hardware \ac{IP} protection include locking-based methods, where designers insert additional gates (controlled by an input key) to thwart reverse engineering~\cite{roy_epic:_2008,rajendran_security_2012,rajendran_logic_2012,LLC,shamsi_ip_2019}. 
When applied at high levels of abstraction, such as \ac{RTL}, these methods are effective because they hide the essential semantics of the design, but entail significant increases in overhead~\cite{pilato_assure_2021}. 
However, there is an ongoing back-and-forth battle between attacks and locking-based defenses, where Oracle-guided attacks~\cite{subramanyan_evaluating_2015,shamsi_icysat_2019,be_sat,cyc_sat,shamsi_kc2_2019} and existence of structural artifacts~\cite{li_piercing_2019,han_does_2021} pose considerable challenges to defenders. 

\acp{eFPGA}, comprising \acp{CLB} containing \acp{LUT}, flip-flops, and routing logic, can be programmed to implement arbitrary functionality. 
This allows a designer to implement ``sensitive'' parts of the design in an \ac{eFPGA}, post-fabrication, by means of setting the configuration bit-stream--- unseen by potentially untrusted parties during manufacture. 
Hence, \ac{eFPGA}-based redaction is a potential panacea for reverse-engineering attacks; the regular structure of an \ac{eFPGA}, avoids apparent structural biases while appearing to pose a challenge to attackers by introducing a large key-space (i.e., configuration bitstream)~\cite{hu_functional_2019,mohan_hardware_2021}.  
Using an \ac{eFPGA} for redaction offers expressiveness and complexity compared to focusing on replacing parts of a design with \acp{LUT}~\cite{kolhe_security_2019} or by  obfuscating routing structures~\cite{kamali_interlock_2020}.

However, how to identify portions of a design to redact is an open issue; the designer must not only identify the ``sensitive'' parts but also decide how much of them can be moved onto the \ac{eFPGA}. This requires  careful evaluation of the security benefits of \ac{eFPGA} implementation while limiting  overhead. 
Prior work studied this problem from a \ac{HLS} perspective, where security is explored in terms of operations redacted and the number of cells on which those operations are mapped~\cite{hu_functional_2019}, or by mapping the logic that differs between variants of the same functionality~\cite{chen_decoy_2020}. 
While \ac{HLS} studies provide insight into targets for redaction, they do not fully characterize practical implications of an \ac{eFPGA} fabric to support redaction; for instance, the \ac{eFPGA} fabric dimensions needed and the fabric interfaces are not apparent. 

``Designer-directed'' redaction at the \acl{RTL} using a custom fabric generation flow~\cite{mohan_hardware_2021} produces a fabric for parts of the design the designer intuits as ``security critical''.  The security analyses show that the fabric offers a promising level of SAT-attack resilience, but the evaluation is limited to a small number of designs. Our study sheds light on practical issues with  \ac{eFPGA} redaction at the \acl{RTL} by extending the analysis of prior work. We adapt an open-source \ac{FPGA} design flow and redact a wider range of \acp{IP}. 

%




\subsection{Open Source (e)FPGA Design Flows}

The trends in heterogeneous computing have increased interest in \acfp{eFPGA} fabrics due to their flexibility and adaptability. 
In commercial products, \acp{FPGA} are tightly integrated to processors in a single-chip, serving as a co-processor or a programmable accelerator \cite{intel_xeon_fpga, PSchiavone_tvlsi_2021}.
Thanks to \acp{eFPGA}, the peak performance of a \ac{SoC} can be improved by 3.4$\times$ along with a 2.9$\times$ power reduction.
Different \acp{SoC} require different \ac{eFPGA} fabrics from architecture to physical layouts, depending on the application context. 
For instance, \acp{eFPGA} designed for machine learning applications require a high density of \textit{Digital Signal Processing} (DSP) blocks, embedded memories, and architectural enhancements which can implement \textit{Multiply-accumulate} (MAC) operations efficiently.

\begin{figure}[t]
    \centering
    \includegraphics[width=\columnwidth]{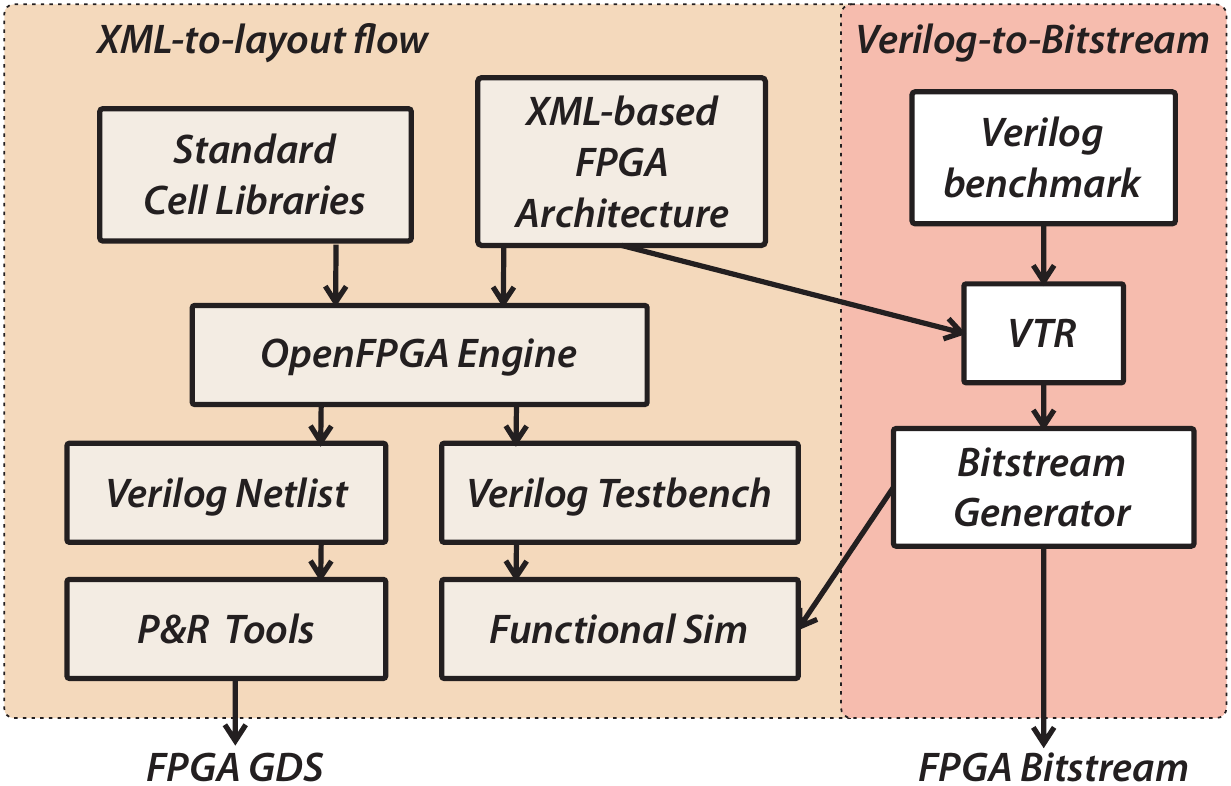}
    \caption{Open-source \ac{eFPGA} design flows: (a) XML-to-layout generation for chip designers; and (b) Verilog-to-Bitstream generation for end-users.}
    \label{fig:openfpga_flow}
\end{figure}

\begin{figure*}[t]
    \centering
    \includegraphics[width=0.9\textwidth]{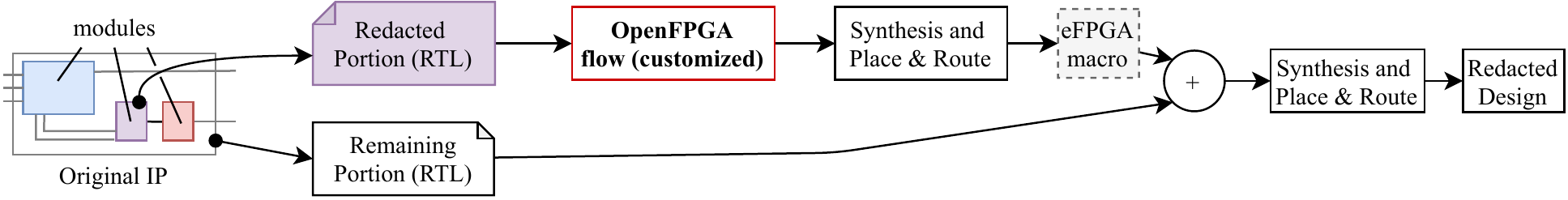}
    \caption{An \ac{eFPGA}-based redaction flow for \ac{RTL} \ac{IP}, where the redacted portion (module) is manually picked by a designer. We adapt the OpenFPGA flow to produce the required eFPGA fabric, which we then treat as a macro to be connected to the remaining portion of the design. }
    \label{fig:redact-flow}
\end{figure*}

Driven by demand, there are a few open-source tools to prototype customizable (e)FPGAs \cite{tang_openfpga_2020, koch_fabulous_2021, ALi_FPGA_2021}.
\autoref{fig:openfpga_flow} illustrates principles of  OpenFPGA framework to prototype customizable \acp{eFPGA}~\cite{tang_openfpga_2020}. In the \textit{XML-to-layout flow}, chip designers can generate fabrication-ready \ac{eFPGA} layouts using well-known XML-based architecture description languages \cite{JLuu_FPGA_2011, XTang_tvlsi_2018}. The architecture description languages allow designers to customize \ac{FPGA} architecture down to circuit elements, supporting standard cells and highly flexible hardware IPs. The core engine converts the architecture description to synthesizable or technology-mapped Verilog netlists that model a complete \ac{eFPGA} fabric. Then, the auto-generated Verilog netlists can be fed into established ASIC design tool suites, especially \textit{Place\&Route} (P\&R) tools, for generating GDSII layouts and performing sign-offs. In addition, self-testing Verilog testbenches can be automatically generated to ease pre- and post-layout verification. The Verilog testbenches can validate the correctness of an \ac{eFPGA} by simulating a complete process in practice, including bitstream downloading and \ac{eFPGA} operation. As argued in prior work, the ability to create custom, small, fabrics can provide a better fit for redaction compared with off-the-shelf  \ac{eFPGA} \ac{IP}~\cite{mohan_hardware_2021}. 

In the \textit{Verilog-to-Bitstream flow}, end-users can implement HDL designs on the \acp{eFPGA}. HDL designs are first synthesized by Yosys \cite{CWolf_yosys_2013} and physically mapped (packed, placed and routed) on the \ac{eFPGA} programmable resources by VPR \cite{KMurray_trets_2020}. The implemented design is translated to a bitstream which is compatible with configuration protocols of \acp{eFPGA}.
%

Open-source efforts aim to overcome two major technical barriers of contemporary \ac{eFPGA} development: (1) the time-consuming physical design process---by leveraging the sophisticated ASIC design tools rather than manual layouts, and (2) the ever increasing design complexity of associated \ac{EDA} tool-chain---by using well-known open-source \ac{FPGA} architecture exploration tools, e.g., VPR~\cite{KMurray_trets_2020}, rather than developing ad hoc, in-house tools.
Previous work has shown that using the design flows in \autoref{fig:openfpga_flow}, the development cycle of a 160k-\ac{LUT} \ac{FPGA} layout is $\sim$24~hours and its performance is competitive against commercial products \cite{GGore_ispd_2021, tang_openfpga_2020}. 
In this paper, we thus investigate the OpenFPGA framework to implement \ac{eFPGA} fabrics \cite{tang_openfpga_2020} for redaction and give insight into our experience.

\section{Redaction and Practical Challenges\label{sec:proposed}}

\subsection{Overview of the Redaction Process}

\autoref{fig:redact-flow} illustrates a general \ac{eFPGA}-based redaction flow for redacting individual, hierarchically designed \acp{IP} at the \acl{RTL}. 
Typically, designers redact a module after designing them~\cite{pilato_assure_2021,mohan_hardware_2021}, which points towards hand-crafted modules in the \ac{IP} as the starting point for potential redaction targets. 
To prepare the redaction fabric, we run the selected module through the OpenFPGA framework~\cite{tang_openfpga_2020} which selects and generates the smallest \ac{eFPGA} fabric configuration given an architecture definition. 
We simulate the generated fabric to verify that the intended functionality is correct, and if so, take the synthesizable Verilog netlist through a physical design flow that comprises synthesis, followed by floorplanning, placement and routing. 
In contrast to related work~\cite{mohan_hardware_2021}, we treat the eFPGA fabric as a \textit{macro}. 
After integrating this macro with the rest of the design, the \ac{IP}, as a whole, is put through the design flow, resulting in a final GDSII file. 

\subsection{Practical Considerations}

There are several practical considerations that designers need to keep in mind when using \acp{eFPGA} for redaction, including the fabric utilization, impact on timing, and the area overhead introduced by integrating a fabric. 

\textit{Resource Utilization}: When one redacts an \ac{IP}, the selected module(s) (the ``redaction modules'') needs to fit into the \ac{eFPGA} fabric; designers need to be aware of the resources available in a particular fabric size, especially if one were to adopt an HLS-based ``top-down'' approach to increase cell usage~\cite{hu_functional_2019}. 
The alternative approach is to find a fabric size that matches the requirements of the ``designer-directed'' redaction choice~\cite{mohan_hardware_2021}. 
However, the minimum fabric size is driven by different factors of the redaction module. The interface of the module (number of inputs and outputs) will affect the number of I/O tiles required, while the number of state elements (registers/flip-flops) will affect the number of \acp{CLB}.  Either factor can dominate the final eFPGA size, causing a sub-par use of the fabric used for redaction.

\textit{Timing}: The chosen redaction unit can possibly lie in the critical path. 
\ac{FPGA} structures tend to have longer delays compared to full ASIC designs due to the general nature of the large pool of available gates for logic and routing. 
Thus, the redacted portion in the \ac{eFPGA} will likely be slower compared to the same design implemented directly in the ASIC.  The designer should be aware of the impact on the overall \ac{IP}'s timing characteristics, including the effect of the redacted portion, otherwise the targeted performance is compromised.

\textit{Area:} In addition to timing issues, the introduction of an \ac{eFPGA} fabric will have considerable implications on area, particularly as the 
number of \ac{CLB} and I/O tiles increases non-linearly with each increase in the square \ac{eFPGA} fabric's dimensions.  
This places another constraint on the design portion selected for the redaction---a redaction choice that requires a fabric that encompasses too much area, in the context of the \ac{IP} as a whole, could be too impractical. 
%
In a related vein, the module selected for redaction could have numerous instances in the \ac{IP}; the designer could possibly create a larger fabric to redact several instances, instantiate multiple \acp{eFPGA}, or possibly choose to redact only one.  

To gain insights into these practical considerations, we explore the fabrics needed to redact different parts of typical IPs (presented in \autoref{sec:results}). On top of these practical considerations, we need to consider the security implications of \ac{eFPGA} fabrics---we explore this in the next section.

\section{Security Analysis of eFPGA Fabrics\label{sec:security}}

For more insight into the security offered by using \ac{eFPGA}-based redaction, we explore and discuss their characteristics, in particular, SAT attack resilience, given the miter-based SAT attack's strength against other locking/obfuscation approaches~\cite{LLC}. 
Previous work has suggested that large \ac{FPGA} bitstream lengths make SAT-based attacks impractical~\cite{hu_functional_2019} and the results in Mohan \textit{et al.}'s evaluation appear to support this claim~\cite{mohan_hardware_2021}. 
In this work, we begin to investigate how the different structural elements of the \ac{eFPGA} contribute to SAT resilience. 
We also perform a security evaluation of different fabric sizes in a high performance computing (HPC) environment, with jobs running on a compute node with an Intel Xeon Platinum 2.9~GHz with 64--512~GiB of RAM. 


\subsection{Threat Model and Assumptions} 
We consider an attack model favoring an attacker with access to a fully-scanned and unlocked design (i.e., an oracle) in addition to the netlist of the \ac{IP}; this is typical from prior work~\cite{mohan_hardware_2021}. 
An adversary has to overcome three challenges before they can launch a reverse-engineering attack. First, they have to isolate the \ac{eFPGA} fabric from the rest of the \ac{IP}; this is possible since the regular structure of the fabric stands out from the rest of the design (as seen in~\autoref{fig:layouts}). 
Second, for oracle access, they should be able to control and observe the inputs and outputs of the fabric and all the flip-flops. As a worst-case assessment, we endow the adversary with these capabilities although there are orthogonal efforts to mitigate this attack model~\cite{limaye_thwarting_2020}. 
Third, the attacker cannot trivially extract the \ac{FPGA} bitstream~\cite{hu_functional_2019}. 
Physical attacks (e.g., optical probing~\cite{rahman_key_2020}) are out of scope. 

\subsection{Why do \acp{FPGA} appear to be SAT-attack resilient?}
Using this threat model and assumptions, the first hurdle for an adversary is to formulate the design and miter circuit inputs to a SAT solver~\cite{subramanyan_evaluating_2015}, treating the configuration bitstream as the ``key''.
SAT solvers fail in the presence of combinational loops (cyclic designs)~\cite{cyc_sat}, producing unstable results or repeatedly returning distinguishing input patterns. 
These loops emerge from the re-configurable routing network of the \ac{eFPGA}. The sequence of re-configurable logic represented by the chain of \acp{LUT}/\acp{CLB} interconnected by this re-configurable network adds a polynomial complexity to the SAT formulation. 


To launch a SAT attack on designs with (potential) loops, like eFPGAs, we need to pre-process the netlist to add constraints to break the loops. Multiple approaches have been proposed to modify the SAT attack for cyclic designs. CycSAT~\cite{cyc_sat} and BeSAT~\cite{be_sat} are two approaches. 
However, \acp{eFPGA} have hard combinational loops what CycSAT cannot resolve. These hard loops are intertwined; even when CycSAT breaks a loop to make the circuit acyclic, atleast one loop remains. The acyclic constraints generated by CycSAT overlook such loops and live-locks the solver into repeating the same \acp{DIP}. Be-SAT can break such loops by pruning the keys leading to live-lock \acp{DIP}. However, it has exponential complexity in  key-size.  
IcySAT~\cite{shamsi_icysat_2019} is a practical loop-breaking alternative that finds a subset of feedback nets that when "removed" make the netlist acyclic. The circuit is then unrolled with respect to these feedback nets, with an unroll factor equalling the size of the feedback set. The unrolled circuit can feed into any SAT attack tool. 

\subsection{Security Evaluation Setup}
To formulate a SAT attack to recover the \ac{eFPGA} bitstream, the \ac{eFPGA} netlist has to be redefined as a key-controlled netlist, with the configuration bitstream being the key. 
In an \ac{eFPGA}, a bitstream is loaded into the configuration \acp{FF} as a sequence of configuration bits. 
The configuration \acp{FF} are interconnected as a scan-chain driven by a  programming clock ($prog\_clk$). 
In our attack setup, we write a script to expose the configuration \acp{FF} as primary key inputs by traversing along the scan chain. 

To identify the configuration scan chain, we do a depth-first search (DFS) starting from the $scan\_in\_head$ port, until we reach the $scan\_in\_tail$. 
All \acp{FF} in the traversal path driven by the programming clock ($prog\_clk$) store the configuration bitstream. 
The order in which the configuration \acp{FF} are detected along the path corresponds to the bitstream order. 
The detected configuration \acp{FF} are exposed as primary key inputs to convert the \ac{eFPGA} netlist into a typical SAT attack netlist. This key-exposed netlist is fed to our implementation of IcySAT~\cite{shamsi_icysat_2019} to unroll the hard loops. 
To model an oracle, we use the same netlist, but add constraints to set the key-bits to the configuration values from the bitstream generated in the OpenFPGA flow. The unrolled netlist and the oracle netlist are  used with KC2 attack tool~\cite{shamsi_kc2_2019} in our experiments.

\subsection{Impact of Fabric Size\label{sec:size-analysis}}

\begin{table}[t]
\caption{The amount of unrolling and the number of clauses when preparing the \ac{eFPGA} fabric for the SAT-based attack}
\label{tab:sat-clauses}
\centering
\begin{tabular}{@{}ccc@{}}
\toprule
Fabric & Unroll Factor & \# Clauses (millions) \\ \midrule
3$\times$3    & 190           & 6                   \\
4$\times$4    & 628           & 67                  \\
5$\times$5    & 1441          & 324                 \\ \bottomrule
\end{tabular}%
\end{table}

To better understand the impact of fabric size, we synthesized square \acp{eFPGA} fabrics of various configurations, based on \acp{CLB} with eight 4-\acp{LUT} (more details in \S\ref{sec:configs}) surrounded by I/O tiles, and converted them into unrolled netlists (as described in the previous section). 
As attack difficulty correlates with how the circuit is modeled for the SAT attack, we categorize \ac{eFPGA} fabrics in terms of number of feedback nodes to be broken (Unroll Factor) and the clause size of \ac{eFPGA} netlist, shown in \autoref{tab:sat-clauses}. 
The size of IcySAT unrolled netlist equals the product of unroll-factor and the number of clauses required to represent the original circuit---both factors contribute to the clauses added per iteration of the SAT attack and contributes to attack complexity. 

Table \ref{tab:attack_results} shows the result of attack on different fabric sizes. FPGA fabrics mapped with multiple designs shown in Table \ref{tab:synthesis-results} was subjected to SAT attack. It was observed that the complexity of the attack increases exponentially as we increase \ac{eFPGA} fabric size. 
Our attack of the 3$\times$3 fabric was successful, and was completed on average in 534~s. 
We tried to launch similar attacks on the 4$\times$4 and 5$\times$5 fabrics, but these were not able to complete within 48~hours, which suggests that at least a 4$\times$4 fabric should be selected, as a minimum, for redaction. 
In attempting to launch these attacks, we needed to increase the amount of RAM available to KC2 (we doubled the allocation in the HPC system each time); the attack on 4$\times$4 fabric only stopped crashing with 128~GiB of RAM, while the attack on the 5$\times$5 fabric required 512~GiB. 
To see if the attack time of a fabric is affected by the design it implements, we tried to attack three different designs in the 3$\times$3 fabric. 
No significant differences were observed in attack time. 
This suggests that, for a fixed fabric, the attack complexity might be independent of the bitstream. In future, we will extend this work to validate the observation with results on larger fabrics.

\subsection{Analysis of Different Parts of the Bitstream \label{sec:category-analysis}}

\begin{table}[t]
\centering
\caption{Results from Attacking Different Parts of the Bistream}
\label{tab:bitstream_analysis}
\begin{tabular}{@{}ccccc@{}}
\toprule
Bitstream    & Clauses & Variables & Time   & Key-size \\ \midrule
I/O bits     & 6197406 & 2436085   & 68.7   & 12           \\
Routing Bits & 5951166 & 2313613   & 105.96 & 336          \\
Logic Bits   & 6043126 & 2359351   & 67.4   & 215         \\ \bottomrule
\end{tabular}
\end{table}

Bits in the \ac{eFPGA} bitstream can be categorized depending on the part of the \ac{eFPGA} they configure. 
We identified three main parts 
\textit{routing configurations bits} comprising Switch Block and Connection Block configuration bits (i.e., the logic within tiles for routing signals), \textit{logic configuration bits} comprising the configuration bits for the \acp{CLB}, and the \textit{I/O configuration bits}, which configures the I/O ports to be input or output in the \ac{eFPGA}. We attempted a ``piece-wise'' analysis of the challenge in recovering the different  configuration bits, assuming all others are known, for insights into which elements of an \ac{eFPGA} might be harder/easier to recover. This can inform redaction-centric \ac{eFPGA} fabric design in future. 

We launched a partial SAT attack on the 3$\times$3 fabric to recover a particular class of configuration bits while assuming the others are unknown. \autoref{tab:bitstream_analysis} shows the recovery time for different components of the bitstream while assuming other bits are known. We observed that routing and logic bits have similar complexity in terms of attack run-time required per bit of bitstream. 
This is anticipated as CLBs and routing units constitute MUX-trees with configuration bits either controlling select inputs of MUXes or acting as a data input of a MUX and hence represent a similar logic at gate-level of abstraction.

However attacking I/O bits resulted in a larger attack run-time per bit. We intuit that this arises from the low output corruptibility resulting from the I/O bits. For partial SAT attack, we assume routing and CLB bits are known; this configures the inputs being used by the \ac{eFPGA}.
For example, assume an \ac{eFPGA} configuration that uses 2 out of 10 possible inputs. Since the \ac{eFPGA} logic is configured to use the 2 inputs, changing the don't-care inputs does not lead to a \ac{DIP}, as required by the SAT attack. The SAT solver has to search for the correct I/O bit configuration such that the two care inputs can be used to find a \ac{DIP}.  This limits \ac{DIP} equivalence class, increasing solver time. 

\subsection{Exploring the Impact of Partial Bitstream Recovery}
\begin{figure}[t]
    \centering
    \includegraphics[width=0.8\columnwidth]{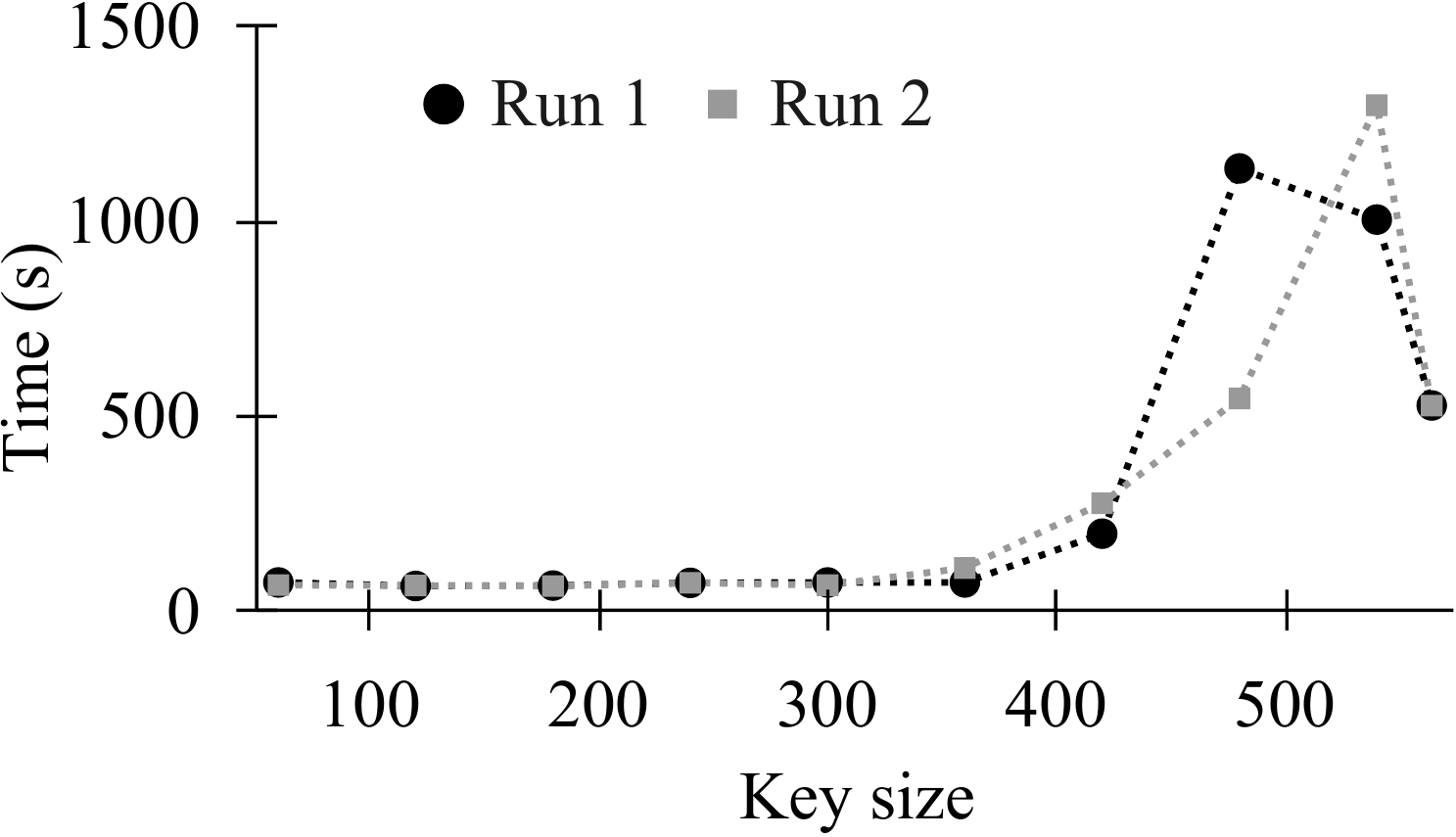}
    \caption{Time dependency on key-size for partial SAT Attack for two random configuration bits chosen to be key-bits, for a 3$\times$3 \ac{eFPGA} fabric.}
    \label{fig:key_vs_time}
\end{figure}

In this section, we explore the security compromise when an attacker can recover a part of bitstream through a side-channel, replicating and extending the study in prior related work~\cite{mohan_hardware_2021}. 
For instance, since the ASIC/eFPGA interface could be identified from the netlist, the attacker might be able to guess the I/O configuration bits, thus reducing the key search space. 
To explore the security under such a scenario, we explore how the attack time varies with number of unknown key-bits. 
\autoref{fig:key_vs_time} shows how the attack-time varies with number of unknown key-bits.

Counter to intuition that attack run-time is proportional to the number of unknown configuration bits, we found that when a small subset of configuration bits are known, the attack-time is greater when compared to the attack-time when all key-bits are unknown. To confirm that this is not a consequence of the random nature of configuration bits chosen to be key-bits, we run two trials choosing different random set of configuration bits. The information on known key-bits is added to the circuit formulation as constraints. Consequently, the effective number of variables remains the same. When there is a substantial information on the key variables, it reduces the search space of the Davis-Putnam-Logemann-Loveland (DPLL) algorithm \cite{dpll} within the solver,  reducing the solver effort. However, when a  small subset of keys are known, this information may burden the solver, without significantly pruning the search space. This may unpredictably increase or decrease the run time as suggested by this experiment. We conclude that the attacker needs to recover a substantial number of key-bits to invalidate the SAT resiliency claim. To verify this argument from a practical perspective, we launched an attack assuming that the attacker was successful in recovering all I/O configuration bits (12/563 bits) in a fabric by snooping at the ASIC-eFPGA interface and launched an attack with 551 unknown keys and 12 known I/O configuration bits. The partial SAT attack took about 1045 seconds when compared to 545 seconds for the full SAT attack.

\begin{table}[t]
\centering
\caption{Results of SAT attack on different fabric netlists}
\label{tab:attack_results}
\resizebox{\columnwidth}{!}{%
\begin{tabular}{@{}C{1cm}C{1.5cm}C{1cm}C{1cm}cccc@{}}
\toprule
Fabric size & Circuit & Unroll factor & Bitstream  & \# Clauses & Time (s) & \# Variables \\ \midrule
3$\times$3         & 2-bit adder      & 190           & 563             & 5775606     & 543.5    & 226668  \\ \hline 
3$\times$3         & 1-bit adder      & 190           & 563             & 5775606     & 529.3    & 226668  \\ \hline
3$\times$3         & 2-bit Multiplier & 190           & 563             & 5775606     & 527.6    & 226668  \\ \hline
4$\times$4         & memory write     & 628           & 1904            & \_          & TO       & \_   \\ \hline
5$\times$5         & zero comparator  & 1441          & 4184            & \_          & TO       & \_   \\ \bottomrule
\end{tabular}
}
\end{table}

\section{Experimental Exploration\label{sec:results}}


\subsection{Experimental Overview\label{sec:case-study}}
To explore practical issues and feasibility of \ac{eFPGA}-based redaction, we consider, as a case study, a set of open-source \acp{IP} that comprise hand-crafted modules. These \acp{IP} reflect various application domains and feature in logic locking/obfuscation literature. For each \ac{IP} we examine the modules and their characteristics. We select several modules to put them through the OpenFPGA flow to identify the fabric size required to redact the module, synthesize the fabric to produce an eFPGA macro, \autoref{fig:macro}, and then put the combined \ac{IP} through the physical design flow \autoref{fig:full}. For this study, we target the FreePDK 45~nm technology library~\cite{FreePDK45}. We do synthesis using Cadence Genus 18.1 and use Cadence Innovus 18.10 for physical design. Synthesis was performed on a server with AMD EPYC 7551 (32 Core, 512 GB RAM). 

\subsection{Case Study IPs}

\begin{table}[b!]
\centering
\caption{Characteristics of the Case Study IPs}
\label{tab:IP-stats}
\begin{tabular}{@{}cccc@{}}
\toprule
\multirow{2}{*}{IP} & \multirow{2}{*}{\# Modules} & \multicolumn{2}{c}{Module Interfaces (Range in Bits)} \\ \cmidrule(l){3-4} 
 &  & Input Bits & Output Bits \\ \midrule
AES & 9 & 10--128 & 8--128 \\
GCD & 8 & 8--45 & 1--18 \\
GPS & 12 & 6--128 & 1--256 \\
PicoSoC & 10 & 8--96 & 4--86 \\
12-bit Adder & 1 & 24 & 13 \\
2-bit adder & 1 & 4 & 3 \\ \bottomrule
\end{tabular}%
\end{table}

For this study, we redact a variety of \ac{RTL} \acp{IP} to gain a broader sense of the implications of \ac{eFPGA}-based redaction. 
\autoref{tab:IP-stats} shows the IPs, the number of unique \ac{RTL} modules, and the ranges of input/output bits in the modules. 
The IPs include small designs, like GCD from the OpenRoad project~\cite{ajayi_toward_2019}, to larger designs, like GPS from the MIT Lincoln Labs Common Evaluation Platform~\cite{CEP}. These IPs perform arithmetic and cryptographic operations and appear as targets for obfuscation in prior work~\cite{pilato_assure_2021}. 
Given the number of modules in each IP at the \acl{RTL}, the designer has numerous options for redaction. 
For each IP, we select a module for redaction, as shown in \autoref{tab:synthesis-results}.

To examine the redaction in the context of an \ac{SoC}, we use PicoSoC~\cite{Pico}, which includes the PicoRV32, a size-optimized RISC-V CPU~\cite{Pico}. 
As the designer has freedom to chose what to redact, we redact a portion of the design that affects the CPU function---
for our experiments, we redact the logic that signals whenever the memory is ready. 


For AES, from CEP~\cite{CEP}, we redact two modules: the module which generates the valid\_out signal (which indicates that the encryption is done and the output is ready to be read) and the rconst value. In AES encryption, we need to generate the key for each round (key\_expansion); this function requires the rconst value for each round. This rconst value can be read from the fabric instead of hard-wiring. These modules are used independently in the AES module. From the CEP, we also protect GPS IP; we redact the ``C/A Code Civilian Acquisition or Access Code'' (CACODE) module. 
Additionally, to understand our study not only on larger IPs but some general purpose \acp{IP}, we redact a 2-bit and 12-bit adder from the GCD IP from OpenRoad project~\cite{ajayi_toward_2019}. 
For the GCD IP, we redact logic that subtracts/compares data to zero. 





\subsection{Using OpenFPGA for eFPGA Generation \label{sec:configs}}
The OpenFPGA flow allows a designer to specify various \ac{FPGA} architectures, for instance, by choosing different \ac{CLB} designs. 
We selected a simple FPGA architecture that has appeared in prior work~\cite{KMurray_trets_2020,tang_openfpga_2020,koch_fabulous_2021,ALi_FPGA_2021} that comprises \acp{CLB} built with eight 4-input LUTs, which we specify in the .xml architecture file for OpenFPGA (as shown in \autoref{fig:openfpga_flow}). 

To produce the \ac{eFPGA}, we replace the ``out-of-the-box'' I/Os pads with simpler input and output pins since the fabric is embedded in an ASIC; in our redactions, the fabric does not need to be able to communicate with the rest of the \ac{SoC} or off-chip. 
This simplification reduces the area overhead tremendously (compared to a non-embedded \ac{FPGA} fabric). 
For connections within the fabric we use 2-input multiplexers whose select line is controlled by the configuration bitstream. 
An added benefit of the OpenFPGA flow~\cite{tang_openfpga_2020} is that it automatically generates testbenches for functional verification as well as the required SDC files to disable combinational loops during synthesis of the fabric. 
The eFPGA fabric is synthesized, placed, and routed separately from the rest of the IP, to produce an eFPGA macro. 






\begin{figure}[t!]
    \centering
    \subfloat[\label{fig:macro}]{{\includegraphics[width=0.5\columnwidth]{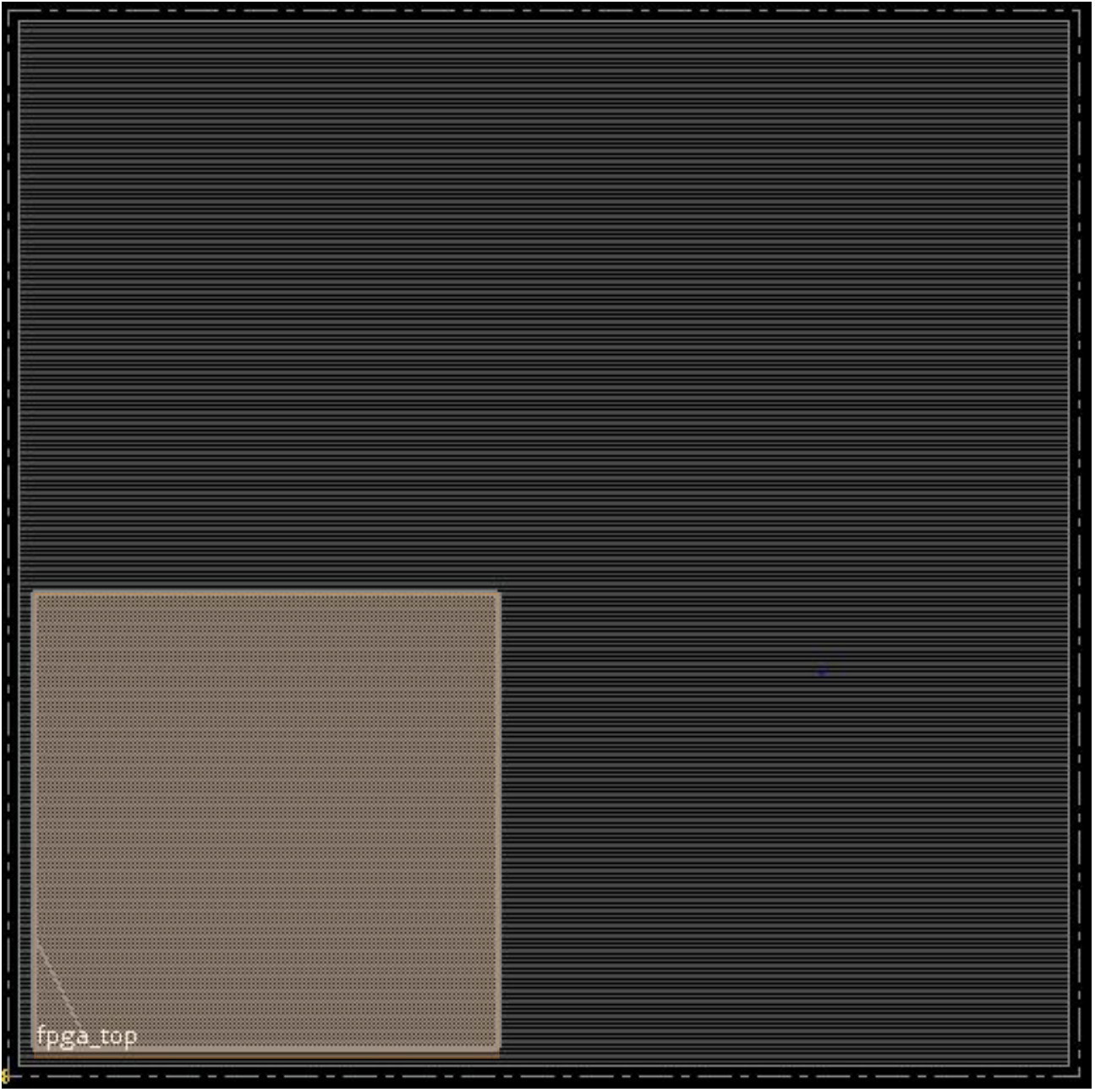}}}%
    \hfill
 \hspace{-1em}
    \subfloat[\label{fig:full}]{{\includegraphics[width=0.5\columnwidth]{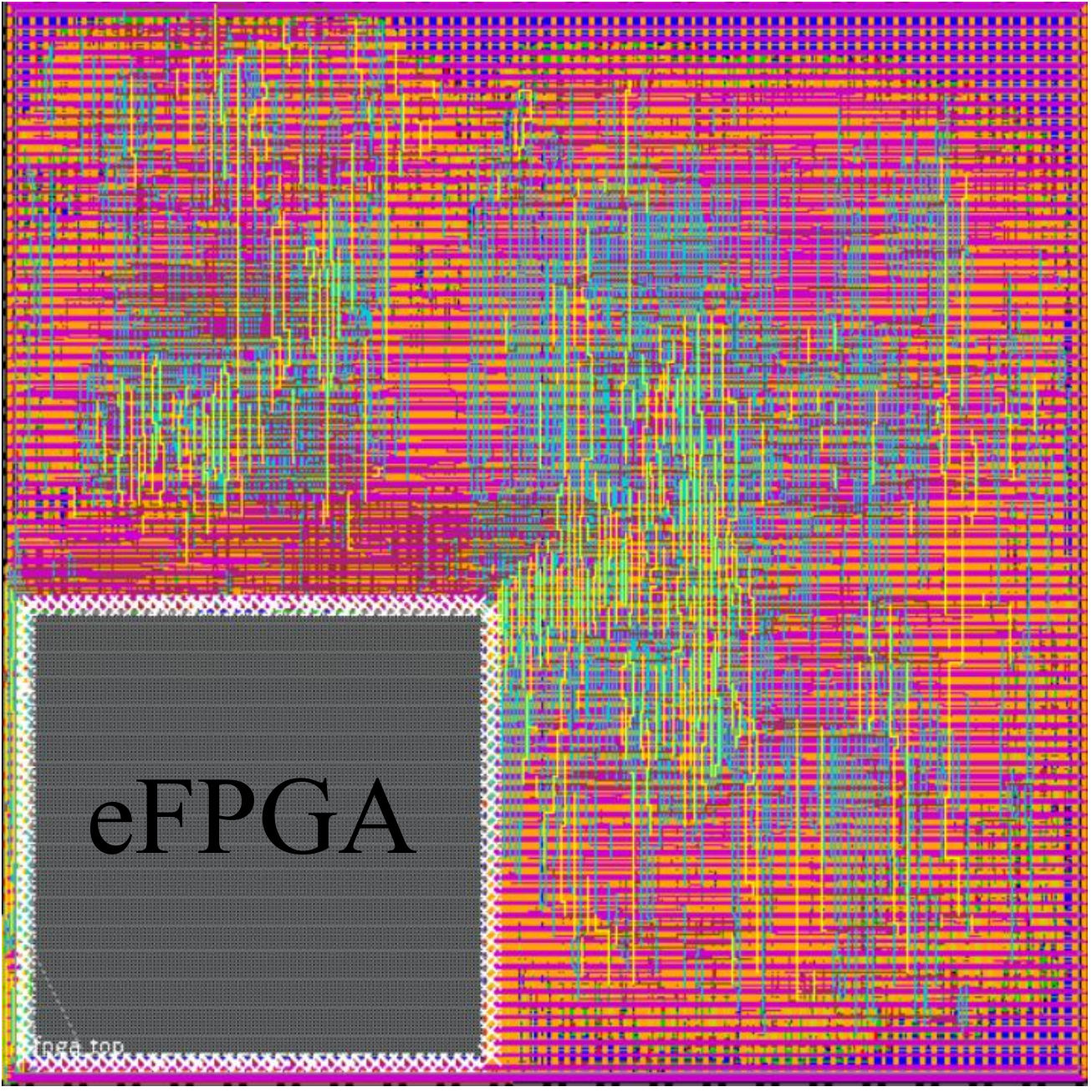}}}%
    \caption{(a) 4$\times$4 eFPGA fabric in the bottom-left corner of the floor plan before placing the rest of the IP. (b)  Complete PicoSoc design with the eFPGA fabric and remaining logic.\label{fig:layouts}} 
    
\end{figure} 

\begin{table*}[t]
\centering
\caption{Characteristics of Redacting Various IPs}
\label{tab:synthesis-results}
\resizebox{\textwidth}{!}{
\begin{tabular}{@{}ccC{1cm}C{1cm}cC{1cm}ccccc@{}}
\toprule
& & \multicolumn{3}{c}{ASIC-only} & \multicolumn{4}{c}{OpenFPGA} & \multicolumn{2}{c}{Redacted IP Area ($\mu m^2$)} \\ \cmidrule(l){3-5} \cmidrule(l){6-9} \cmidrule(l){10-11}
IP & Module  & Critical Path? & Module area in ASIC & Area ($\mu m^2$) & Fabric & I/O utilization (\%)& Resource use (\%) & Bitstream & eFPGA Portion & Total \\ \midrule
GPS & cacode  & No & 296.1 & 273806.0 & 6$\times$6 & 21 & 81 & 9237 & 102312.1 & 570964.5 \\\midrule
\multirow{3}{*}{ GCD } & zero\_comparator  & No & 7.1 & 403.2 & 5$\times$5 & 71 & 25 & 4184 & 45872.5 & 68531.4 \\
& mux & No & 29.8 & 403.2 & 8x8 & 68 & 6 & 16341 & 185230.4 & 264534.5 \\
& subtractor  & No & 74.5 & 403.2 & 8x8 & 80 & 20 & 16341 & 185230.4 & 264614.3 \\ \midrule
AES & valid\_output/rconst  & No & 84.4 & 283944.9 & 6$\times$6 & 67 & 89 & 9237 & 102312.1 & 562648.1 \\\midrule
PicoSoC & memory\_write  & No & 259.0 & 82705.8 & 4$\times$4 & 75 & 100 & 1954 & 21182.4 & 138115.6 \\\midrule
\multirow{2}{*}{ ADDER } & 12-bit Adder  & No & 227.8 & 227.8 & 7x7 & 71 & 19 & 11164 & 128535.2 & 128535.2 \\
& 2-bit adder  & No & 12.1 & 12.1 & 3$\times$3 & 58 & 100 & 564 & 6207.1 & 6207.1 \\ \bottomrule
\end{tabular} }
\end{table*}

\begin{table}[t]
\centering
\caption{Comparison of Area, Power and Delay overhead with integration of different fabric sizes in PicoSoC}
\label{tab:differentfabric-results}
\begin{tabular}{@{}cccc@{}}
\toprule
Design & Area ($\mu m^2$) & Power (mW) & Delay (ns) \\ \midrule
PicoSoC & 108307 & 30.0 & 1.284 \\
PicoSoC + 3$\times$3 & 116316 & 40.2 & 1.878 \\
PicoSoC + 4$\times$4 & 137330 & 49.3 & 2.261 \\
PicoSoC + 5$\times$5 & 173725 & 56.3 & 3.860 \\
PicoSoC + 6$\times$6 & 259508 & 68.7 & 4.708 \\\bottomrule
\end{tabular}
\end{table}

\subsection{Redaction Results}

\begin{figure}[b!]
    \centering
    \includegraphics[width=0.9\columnwidth]{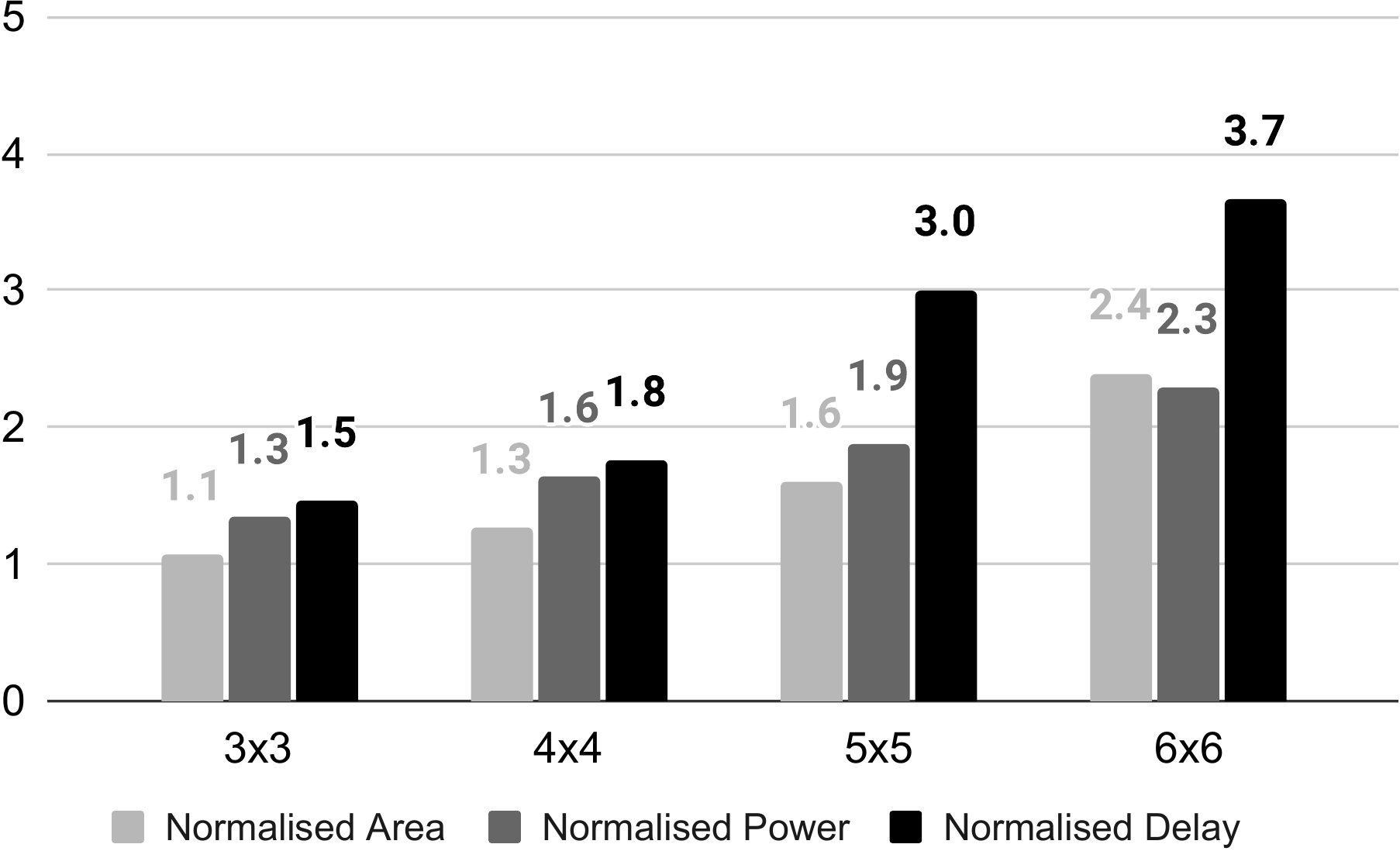}
    \caption{Comparison of area, delay and power overhead of integration of different fabric size with the original design in PicoSoC. Area normalized to 0.11mm$^{2}$, Power normalized to 30~mW and Delay normalized to 1.284~ns}
    \label{fig:norm-fabrics}
\end{figure}

For each IP, we redact a module by replacing its instantiation in the original IP's \ac{RTL} with the fabric macro generated using OpenFPGA, after simulating the eFPGA fabric to ensure that the redacted functionality is implemented correctly. 
The required fabric size for implementing the redacted module ranges from 4$\times$4 to 8$\times$8. 
The overall design is then synthesized, placed, and routed as shown in \autoref{fig:full}. 
For comparison, we also synthesized, placed, and routed each \ac{IP} without redaction. 
The post-synthesis area results are shown in \autoref{tab:synthesis-results}, from reports produced by Cadence Innovus. 

\subsubsection{Area}
\autoref{tab:synthesis-results} shows that there is a significant amount of area overhead associated with the redaction method in general. 
This places a burden on the designer to properly select the best module(s) to achieve their security level with a reasonable overhead. 
Depending upon the size of the original design, the impact can be relatively characterised as practically possible or not feasible at the allocated budget. 


To better understand the area overhead, we take the PicoSoC IP and randomly pick a module to redact such that the fabric required will have different sizes. The result of this exploration is shown in \autoref{tab:differentfabric-results}. Area increases in the range from 10\% to 140\% suggesting that redacting a function and then integrating it with ASIC is not a simple addition of two IPs. 
It requires different placement, floorplan, and routing of the design due to the constraints added by the addition of eFPGA and its timing requirement. Thus the area increases as a non-linear function of fabric sizes (\autoref{fig:norm-fabrics}). 

Moreover, during our experiments, we found that some modules require larger fabrics due to the number of input and outputs of that module. 
This forces us to increase the size of the fabric and impacts the resource utilization in the fabric. \autoref{tab:synthesis-results} points out this issue; consider the 12-bit Adder, where due to limited number of I/Os the fabric needs to be expanded up to 7$\times$7, resulting in only 19\% CLB usage. In other cases, utilisation is better---redaction of AES~\cite{CEP} module, suggests more efficient fabric utilisation (67\% I/O and 89\% CLBs) which is quite acceptable from the designer perspective. 

\subsubsection{Power} 
\autoref{tab:differentfabric-results} shows the variation in the power consumption as FPGA fabric varies, as reported by Cadence Innovus. When we compare the power consumption of the module to be redacted in ASIC implementation with the same function mapped to an eFPGA fabric; there is an increase in the expected power consumption because of the extra switching gates and routing multiplexers to connect the fabric, resulting in 30\%-130\% increases over the original design. 
In terms of power consumption, there is an inevitably large penalty, even for small fabrics. 

\subsubsection{Delay} 
For better understanding the impact of fabric sizes on delay of the overall \ac{IP}, we perform a similar experiment as for area \autoref{tab:differentfabric-results}. Delay has a more prominent impact compared to the impact on area and power, i.e., 50\% to 270\% increase. 
This is apparent from the FPGA architecture, where we need more routing channels to connect tiles to every other tile in a fabric. This contributes to the additional delay. \autoref{fig:norm-fabrics} illustrates a comparison of normalised delay with for the different fabric sizes. 


\section{Discussion\label{sec:discussion} and Future Outlook}

In this work, we explored the feasibility and other practical issues of eFPGA-based redaction. 
We performed a security analysis to investigate the characteristics of eFPGAs that contribute towards its SAT attack resilience, and characterized the impact of redacting different modules in a variety of IPs at the register transfer level.

As discussed in \autoref{sec:case-study}, we explored the redaction of a variety of \acp{IP} for insights into the practicality of \ac{eFPGA}-based redaction. 
Generally, for bigger designs like GPS, AES, and PicoSoC, the overhead introduced by integrating the eFPGA is feasible. 
However, for smaller IPs, the approach is not feasible; for instance, GCD's total design area is smaller than the smallest available fabric (shown in \autoref{tab:synthesis-results}, resulting in a drastic increase in area. 
Thus, framing \ac{eFPGA}-based redaction as a general IP-level protection mechanism might not be practical, despite the high-level SAT-attack resilience.

With eFPGA-based redaction, we are ``overallocating'' resources for redaction (and increasing an attacker's uncertainty).
In future, one may explore feasibility of \ac{eFPGA} redaction for multiple IPs at the \acl{RTL}, i.e., where modules from different IPs share the same redaction fabric. Clearly, there is a complex interplay between fabric size and interface width, fabric utilization by the redacted module, and impact on ASIC quality-of-results, which entails the need for an automated approach to assist with redaction decisions. 

In some respects, our case study emulates a ``designer-directed'' or module-driven approach, in that the module to be redacted is first selected to decide the fabric.  During our experiments, we observed that even though a module is not in a critical path for a design, but after integrating it as an eFPGA with rest of the design, it can dominate the timing.
The full design flow without any prior understanding of the impact can be very time-consuming---for example, it requires us $\sim$8hr for PicoSoC with 6$\times$6 fabric---this suggests that we need a good way to predict the downstream impact of redaction decisions.

Our security analysis (\autoref{sec:size-analysis}) suggests that the attack time depends on the fabric and does not depend on the component redacted, at least in the context of our threat model.
This suggests that a wider variety of heuristics could be considered in deciding what to redact. For example, if a designer is redacting with the intent to ``corrupt'' the output for an unauthorized user without the correct bitstream, identifying what to redact based on the ``highest value'' portion of the design (as suggested by Chen \textit{et al.}~\cite{chen_decoy_2020}) could instead focus on the part of the design with the most impact on the outputs (e.g., identified perhaps through fault analysis). Exploring these alternative approaches to redaction is our future work. 

Finally, our study focused on a single eFPGA architecture and we expect the findings to apply for other eFPGA implementations. 
However, it is possible to vary the complexity of the fabric in terms of blocks and routing, which affects the area, timing, and power characteristics. Our preliminary results in \autoref{sec:category-analysis} point to the fact that different parts of the eFPGA bitstream potentially impact the attack difficulty in different ways---in future, we will explore the possibility of tailoring eFPGA architectures for redaction.

\section*{Resources}
Data for the study in this paper can be found at \cite{noauthor_redaction_2021}.


\IEEEtriggercmd{\balance}
\IEEEtriggeratref{20}

\bibliographystyle{IEEEtran}
\bibliography{references}

\end{document}